\documentclass{elsart}

\usepackage{natbib}
\usepackage{epsfig}
\usepackage{amssymb}

\begin{document}

\begin{frontmatter}

\title{A Review of the Distance and Structure of the Large Magellanic Cloud}

\author{David R. Alves}

\address{Columbia Astrophysics Laboratory, 550 W. 120th St.,
New York, NY, USA; Email: alves@astro.columbia.edu}

\begin{abstract}
The average of 14 recent measurements of the 
distance to the Large Magellanic Cloud (LMC)
implies a true modulus of 18.50 $\pm$ 0.02 mag,
and demonstrates a trend 
in the past 2 years of convergence
toward a standard value.
The distance indicators reviewed are the red clump, 
the tip of the red giant branch,
Cepheid, RR Lyrae, and Mira variable stars, 
cluster main-sequence fitting, supernova 1987A, and
eclipsing binaries.
The eclipsing binaries yield a consistent distance on average;
however, the internal scatter is twice as large as the
average measurement error.
I discuss parameters of 
LMC structure that
pertain to distance indicators, and 
speculate that warps discovered 
using the color of the clump
are not really warps.
\end{abstract}


\end{frontmatter}


\section{Introduction}

The debate about the distance to the LMC has an epic
history full of controversial and dramatic claims
(Walker 2003),
and yet in recent years a standard distance modulus has emerged
due primarily to the
completion of the {\it Hubble Space Telescope} ({\it HST\,})
key project to measure the Hubble constant
(Freedman et al.~2001).
The standard modulus, $\mu_{0} = 18.5 \pm 0.1$ mag,
yields $H_0 = 71\,\pm\,10$
km s$^{-1}$ Mpc$^{-1}$ (total error)
in excellent agreement with that
derived from the {\it Wilkinson Microwave Anisotropy Probe}:
$H_0 = 72\,\pm\,5$ (Spergel et al.~2003),
which lends considerable support to its accuracy.
Moreover, 
it is a recent trend in the literature
that most new LMC distance measurements
indicate $\mu_{0}$ = 18.5 mag,
and that many 
systematic errors in prior studies
are being found and corrected.
In order to illustrate the trend,
this review is restricted to mostly refereed journal papers
published since January, 2002.

\section{Red Clump}

Alves et al.~(2002) presented and
analyzed the first near-infrared data for
LMC red clump giants.  The $K$-band data were obtained 
with the SOFI infrared imager
at the New Technology Telescope.
Additional $V$ and $I$ data were obtained
with WFPC2 onboard the {\it HST\,}.
A comparison of multiwavelength luminosity functions
of the LMC red clump with those of the {\it Hipparcos\,} red clump 
yielded simultaneously the 
mean interstellar reddening correction
and true LMC distance modulus: 
$\mu_0 = 18.493\,\pm\,0.033_r\,\pm\,0.03_s$
(random and systematic error, respectively).
Pietrzy\'{n}ski \& Gieren (2002) also used $K$-band red clump
data to derive 
$\mu_0 = 18.471\,\pm\,0.008_r\,\pm\,0.045_s$,
as did Sarajedini et al.~(2002) 
who obtained $\mu_0 = 18.54\,\pm\,0.10$.
The reddening correction accounts
for the 0.02 mag discrepancy between the results of
Alves et al.~(2002) and Pietrzy\'{n}ski \& Gieren (2002).
Sarajedini et al.~(2002) 
include the systematic errors associated with
the red clump population correction
and the geometric tilt of the LMC in their reported error bar.
Despite these minor differences,
the {\it Hipparcos}-calibrated red clump distance
to the LMC has probably converged.
It is encouraging
that 3 independent estimates of the apparent $K$ magnitude
of the LMC red clump all agree to within an accuracy of 1--2\%.

\subsection{Tip of Red Giant Branch}

Figure 1 shows the 
tip of the red giant branch (TRGB) in the LMC.
These are unpublished {\it HST}/WFPC2 color-magnitude data 
for 17 different fields on lines-of-sight to LMC microlensing events.
I have applied the average reddening correction derived from multiband
red clump data for 6 of the fields
(Alves et al.~2002).
The TRGB calibration from Bellazini et al.~(2001) is shown
assuming $\mu_0 = 18.5$ mag.  The weak metallicity dependence of
the calibration is expressed in terms of color.
Note that the colors of the LMC TRGB stars suggest
a range of metallicities from
[Fe/H] = $-2$ to $-1$ dex with a mode of about $-1.4$ dex.
This comparison shows that the TRGB and red clump methods yield
superb agreement with the standard LMC distance modulus.

\begin{figure}[t]
\begin{center}
\epsfig{figure=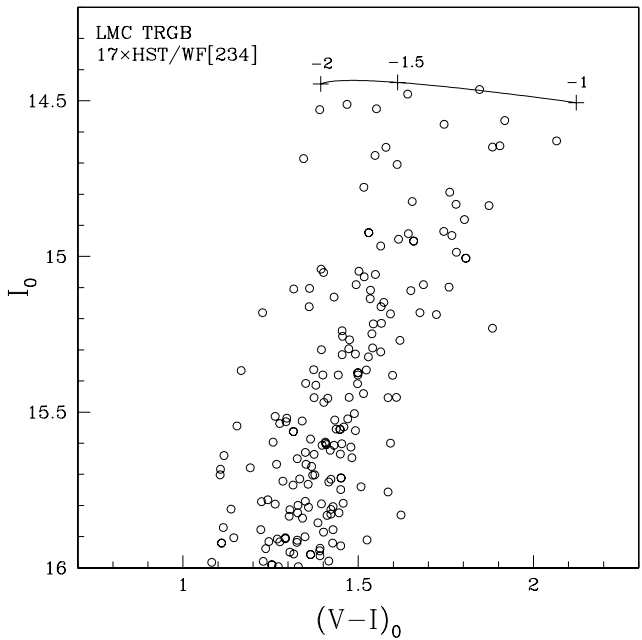,width=3.5in,height=3.5in}
\end{center}
\caption{Color-magnitude array of {\it HST}/WFPC2 data
where LMC TRGB stars (circles) are compared with the
Galactic TRGB calibration (line) assuming $\mu_0$ = 18.5 mag.  
The tick marks along the TRGB calibration label [Fe/H].}
\vskip0.3cm
\end{figure}

\section{Cepheid Variable Stars}

Recent distance estimates
based on Cepheid variable stars in the LMC 
have made use of the densely sampled light curves available
from microlensing surveys like MACHO and OGLE.
Theoretical pulsation models 
reproduce the detailed
shapes of period-folded light curves, and
best-fit models yield the fundamental parameters of each
Cepheid, like luminosity and effective temperature, and hence
the distance.  The Cepheids used for this 
analysis are 
those which display a secondary luminosity maximum 
each pulsation cycle.
These so-called ``bump'' Cepheids 
collectively represent the Hertzsprung Progression
because the bumps occur at earlier phases 
in the pulsation
cycle as period increases.
Bono et al.~(2002) modeled the OGLE $I$-band light curves
for 2 bump Cepheids and 
found $\mu_0 = 18.48$ mag.
Keller \& Wood (2002) modeled the
MACHO $V$ and $R$ light curves for 20 bump Cepheids
to derive $\mu_0 = 18.55 \pm 0.02_r$.
Both studies also found
a level of core overshoot
in good agreement with the calibration from
eclipsing binaries (Ribas et al., these proceedings).

The use of pulsational models of bump Cepheids to derive
the LMC distance is a promising method.  However, 
only a small fraction of the available data
have been used so far.  I have identified 183 bump Cepheids
with periods of 5 to 25 days in the MACHO 
database\footnote{{\tt http://wwwmacho.mcmaster.ca/Data/MachoData.html}},
and out of these 117 are also found in the OGLE 
database\footnote{{\tt http://bulge.princeton.edu/\~\,ogle }}, and
89 in the catalog of four-color light curves
published by Sebo et al.~(2002).  
Furthermore, 153 of these bump Cepheids
are found in the 2MASS database\footnote{
{\tt http://www.ipac.caltech.edu/2mass/}}, which
provides partial light curve coverage in the near-infrared.  
Models will eventually be constructed for
all of the LMC bump Cepheids, and 
parameter estimation will employ more
multiwavelength data.

Table~1 lists
the average differences between the (intensity-weighted)
mean magnitudes published by
OGLE (Udalski et al.~1999) 
and Sebo et al.~(2002) relative to the MACHO
data for bump Cepheids.
The implication is a $V$-band offset
of $-0.032$~mag
between Sebo et al.~(2002) and OGLE;
however, Sebo et al.~(2002) report a
+0.04 mag offset.
Hence the different comparisons of the same two datasets
yield different results.  
I note that their average
is weighted differently and
their comparison is only for Cepheids with periods greater than 10 days.
See Bono et al.~(2002b) for analysis of LMC 
and theoretical
period-luminosity (PL) calibrations and their metallicity dependence.

\begin{table}
\caption{Comparison of Standardized Photometry for LMC Bump Cepheids }
\begin{tabular}{p{1.8in}p{1.8in}p{1.4in}}
\hline
 MACHO -- Other & Ave. Difference (mag) & No. Stars \\
\hline
 $V \ - \ V_{\rm OGLE}$ & $-0.060$ & 117 \\
 $V \ - \ V_{\rm Sebo}$ & $-0.028$ & 89 \\
 $(V-R) \ - \ (V-R)_{\rm Sebo}$ & +0.007 & 89 \\
\hline
\hline
\end{tabular}
\vskip0.4cm
\end{table}

The {\it HST\,} key project calculates true distance modulus using
$\mu_0 = 2.48 \mu_I - 1.48 \mu_V - 0.2 \Delta$[Fe/H],
where $\mu_I$ and $\mu_V$ are apparent moduli 
relative to LMC PL calibrations
(Freedman et al.~2001).  
As a check, consider the following 
application of this method to the Cepheids in the Galaxy.
In this case, however, the Galactic Cepheids are represented 
by only one star, $\delta$ Cephei.
Benedict et al.~(2002)
recalibrated the {\it Hipparcos} parallax 
of $\delta$ Cephei 
using {\it HST\,}/FGS3 and found
$M_V = -3.47\,\pm\,0.10$ and $M_I = -4.14\,\pm\,0.10$~mag.
Adopting the Sebo et al.~(2002)
PL calibrations and 
$\Delta$[Fe/H] = 0.4 dex
yields $\mu_0 = 18.54 \pm 0.29$ mag.  
See Alves (2003) for a review
of the metallicity slope.

\section{RR Lyrae Variable Stars}

Clementini et al.~(2003; C03) presented new $B$ and $V$ photometry
and spectroscopy of RR~Lyrae variable stars in two LMC fields.
They derived an average metallicity of
[Fe/H] = $-1.48\,\pm\,0.03$ dex in good agreement 
with other estimates,
and a luminosity--metallicity slope 
of $0.21\,\pm\,0.05$~mag/dex (in the sense
that the most metal-poor RR Lyrae stars are brightest).
C03 obtained a reddening-corrected magnitude of 
$V_0 = 19.06\,\pm\,0.06$ mag.
Benedict et al.~(2002b) used {\it HST\,}/FGS3 to measure the
parallax of RR Lyrae itself ([Fe/H]=$-1.39$ dex) and
derived $M_V = 0.61\,\pm\,0.11$ mag.
Di\,Fabrizio et al.~(2003) made a Baade-Wesselink analysis of
CM~Leonis ([Fe/H]=$-1.91$ dex) and obtained
$M_V = 0.47\,\pm\,0.04$ mag.
Adopting C03's estimate of 
the metallicity slope,  the weighted average
calibration is $M_V = 0.56\,\pm\,0.04$ mag at [Fe/H] = $-$1.48 dex.
The C03 value for $V_0$ thus implies
$\mu_0 = 18.50\,\pm\,0.07$~mag.

The most recent use of MACHO data 
to derive the LMC distance is based on
first-overtone RR Lyrae stars (Alcock et al.~2004).
Including corrections for blending and crowding bias
based on artificial star tests,
the MACHO data yield $\mu_0 = 18.43\,\pm\,0.06_r\,\pm\,0.16_s$~mag
(C.~Clement \& A.~Muzzin, {\it priv.~comm.\,}).
C03 compared their RR~Lyrae $V$ data
with MACHO data and found an average of difference of
$-0.17$ and $-0.01$ mag in their fields A and B, respectively
(their magnitudes minus MACHO).  
However, Alcock et al.~(2004) found only
7 RR Lyrae stars in common with C03 and these stars yielded
an offset of +0.07 mag (C03 being fainter).
Applying the crowding/blending correction 
yields an offset of $-0.03$ mag, which is consistent within
calibration uncertainties.

Dall'Ora et al.~(2003) recently obtained the first $K$-band data for LMC
RR Lyrae stars in the Reticulum cluster and derived
$\mu_0 = 18.55\,\pm\,0.07$~mag.  
A subsequent analysis of these data
yields a consistent but slightly lower distance modulus
(G. Bono, {\it priv.~comm.\,}).
Finally, additional $K$-band data for field RR Lyrae stars near the
LMC center yields $\mu_0 = 18.48\,\pm\,0.08$~mag
(Borrisova et al.~2004; J. Borissova, {\it priv.~comm.\,}).

\section{Other Distance Indicators }

\subsection{Mira Variables}

Recent work on Mira variables includes a revised zero-point
for the PL calibration based on Galactic globular clusters
(Feast et al.~2002), and a
new {\it Hipparcos} zero-point for thin disk Miras
(Knapp et al.~2003).  Feast (2003) suggests a best-average
PL calibration that leads to 
$\mu_0 = 18.48\,\pm\,0.10$~mag.

\subsection{Cluster Main-Sequence Fitting}

Kerber et al.~(2002) used {\it HST}/WFPC2 data for the LMC
cluster NGC\,1831 and main-sequence models to derive 
$\mu_0 = 18.5$ to 18.7~mag.
Salaris et al.~(2003) used {\it HST}/WFPC2 data for the LMC
cluster NGC\,1866 and a {\it Hipparcos} subdwarf calibration
to derive $\mu_0 = 18.33\,\pm\,0.08$~mag.
However, Groenewegen \& Salaris (2003) subsequently
revised the main-sequence-fitted modulus of NGC\,1866 to
$\mu_0 = 18.58\,\pm\,0.08$~mag.  In this revision, the
reddening correction was derived from
the colors of the cluster's Cepheids.
The field red clump stars around NGC\,1866
are apparently in front of the cluster.

\subsection{Supernova 1987A}

Mitchell et al.~(2002) use the ``Spectral-fitting Expanding
Photosphere Method'' to derive $\mu_0 = 18.46\,\pm\,0.12$~mag
from observations of Supernova 1987A.  They also review 
previous distance measurements based on this object.

\subsection{Eclipsing Binaries }

I refer the reader to 
other papers presented at this meeting for 
details about measuring distances with eclipsing binaries.
In the past 2 years,
Fitzpatrick et al.~(2002) reported a modulus of 
$\mu_0 = 18.51\,\pm\,0.05$~mag for HV\,982,
Ribas et al.~(2002) found 
$\mu_0 = 18.38\,\pm\,0.08$~mag for EROS\,1044,
Fitzpatrick et al.~(2003) found 
$\mu_0 = 18.18\,\pm\,0.09$~mag for HV\,5936, and
Clausen et al.~(2003) found
$\mu_0 = 18.63\,\pm\,0.08$~mag for HV\,982.
The weighted average is 
$\mu_0 = 18.46\,\pm\,0.08$~mag in good agreement
with the standard modulus.
However, the standard deviation 
is about twice as large as the average reported error bar.
Possibly the error bars are too optimistic.
I speculate that an unknown systematic error may 
affect the eclipsing binary distance results.

\section{Summary of LMC Distance}

Table~2 summarizes
the different measurements of the LMC distance published since
January of 2002.
It is customary to calculate an average final result in a review like this one,
and in order to do this I have to make some assumptions about the error bars.
Where both random and systematic errors are given, I adopt
their (arithmetic) sum as the total error, 
and I use this to weight the average.  
Two of the Cepheid-based distance results have incomplete
error estimates, and for these
I adopt $\pm$0.1~mag (see notes in 3rd column).   The result
from Kerber et al.~(2002) is excluded.
The weighted average of 14 measurements 
is $\mu_0 = 18.50 \pm 0.02$~mag
(standard deviation = 0.04~mag).
The reduced chi-squared is less than one, which
suggests that the adopted error bars may be too conservative.

{\it A great American sports journalist once said famously, 
 ``The opera ain't over till the fat lady sings,'' 
to make a point that the outcome of a series of games was not yet 
determined.  Regarding the convergence of published LMC distance 
results, I suggest to you that the fat lady has begun to sing. }

Table~2 demonstrates a remarkably high level
of consistency.
The possibility
that a consensus on the LMC distance 
has been reached seems much more
plausible now than it did when Freedman et al.~(2001) 
reviewed the literature 
at the conclusion of the {\it HST} key project only two years ago. 
See also Walker (2003) for a recent review of distances to
Local Group galaxies.

\begin{table}
\caption{LMC Distance Moduli from 2002 -- Present }
\begin{tabular}{p{1.2in}p{1.65in}p{2.1in}}
\hline
 Method & $\mu_0$ & Reference \\
\hline
Red Clump & $18.493\,\pm\,0.033_r\,\pm\,0.03_s$ & Alves et al.~(2002) \\
Red Clump & $18.471\,\pm\,0.008_r\,\pm\,0.045_s$  & Pietrzy\'{n}ski \& Gieren (2002) \\
Red Clump & $18.54\,\pm\,0.10$ & Sarajedini et al.~(2002) \\
Cepheid & 18.48 & Bono et al.~(2002); $\pm 0.1$ \\
Cepheid & $18.55\,\pm\,0.02_r$ & Keller\,\&\,Wood (2002); $\pm 0.1$ \\
Cepheid & $18.54\,\pm\,0.29$ & Benedict et al.~(2002) \\
RR Lyrae & $18.50\,\pm\,0.07$ & Clementini et al.~(2003) \\
RR Lyrae & $18.43\,\pm\,0.06_r\,\pm\,0.16_s$ & Alcock et al., {\it preprint} \\
RR Lyrae & $18.55\,\pm\,0.07$ & Dall'Ora et al.~(2003) \\
RR Lyrae & $18.48\,\pm\,0.08$ & Borissova et al., {\it preprint} \\
Mira & $18.48\,\pm\,0.08$ & Feast (2003) \\
Main Sequence & 18.5 --- 18.7 & Kerber et al.~(2002); exclude \\
Main Sequence & $18.58\,\pm\,0.08$ & Groenewegen \& Salaris (2003) \\
SN 1987A & $18.46\,\pm\,0.12$ & Mitchell et al.~(2002) \\
Eclipsing Binaries & $18.46\,\pm\,0.08$ & see \S5.4 \\
{\bf Weighted Ave.} & {\bf 18.50$\,\pm\,$0.02} & {\bf (s.d.= 0.04) } \\
\hline
\hline
\end{tabular}
\vskip0.4cm
\end{table}

\section{Structure of the LMC}

In the review just presented, 
the corrections employed by different authors
to account for the inclination
of the LMC in the plane of the sky
were not always self-consistent.
There is no standard model for LMC structure, at least not yet.
Moreover, it is not known that the Population II distance
indicators like RR Lyrae variables are distributed in the
same way as the young and intermediate-age stellar populations
which represent the bulk of the LMC disk, and hence whether or not
a correction appropriate for the disk should apply.
The eclipsing binaries studied in the LMC so far have been mostly
Population~I stars.

In a series of recent papers 
(e.g., van der Marel \& Cioni 2001,
van der Marel et al.~2002),
a new understanding of the structure of the LMC has emerged.
It is now known that the photometric major axis is not
the same as the line-of-nodes.  The LMC is intrinsically elongated.
The line-of-nodes derived from 
a variety of different distance indicators 
that uniformly cover the entire face of the LMC
is $ \Theta = 122.^{\circ}5\,\pm\,8.^{\circ}3$
measured East of North
(van der Marel \& Cioni 2001), while
the radial velocity field of carbon stars implies
$\Theta = 129.^{\circ}6\,\pm\,6.^{\circ}0$ in good agreement
(van der Marel et al.~2002).
The photometric major axis is about 50$^{\circ}$ away.
The inclination is
$34.^{\circ}7\,\pm\,6.^{\circ}2$.
Several independent estimates for the center of the LMC 
are consistent with the optical center.  The notable exception is
the center derived from the velocity field of the HI gas, which is 
disturbed and probably not useful for making inferences about 
the LMC's structure
(van der Marel et al.~2002).

The LMC disk is thick (van der Marel et al.~2002). 
The ratio of rotational to pressure support
is $V / \sigma \sim 2.9\,\pm\,0.9$, where
for comparison the Galactic thin and thick disks have
ratios of 9.8 and 3.9, respectively.
The ratio of vertical to radial disk scale heights is about 1/3
for both the LMC disk and the Galactic thick disk.
The LMC disk is also flared (Alves \& Nelson 2000).
The vertical exponential scale height of carbon stars attains
a value of about 1~kpc at locations 5-6 degrees from center,
and is about 100 pc near the center
(van der Marel et al.~2002).
These inferences about the LMC disk are
based on stellar kinematics and simple equilibrium models.
High precision distance indicators like Cepheids
have not yet resolved the thickness of the LMC disk.
However, observations of the light echoes from Supernova 1987A 
are consistent with a thickness of 
about 1~kpc (A. Crotts, {\it priv.~comm.\,}).

Minniti et al.~(2003) recently obtained 
kinematic data for the metal-poor
stellar population in the LMC using the
European Southern Obervatory's {\it Very Large Telescope}.
The radial velocities derived from the spectra of 43 RR Lyrae stars 
imply a line-of-sight $\sigma = 53\,\pm\,10$~km~s$^{-1}$,
which is larger than that of any other population in the LMC.
For comparison, the carbon stars analyzed by van der Marel et al.~(2002)
have $\sigma = 20.2\,\pm\,0.5$~km~s$^{-1}$.
Minniti et al.~(2003) interpret the RR Lyrae velocity data as evidence for
a stellar halo.  
Therefore, the LMC is the nearest example of a
disk galaxy that has a halo, but not a bulge, and in this 
way it is similar to M33 (Gebhardt et al.~2001).

\begin{figure}[t]
\begin{center}
\epsfig{figure=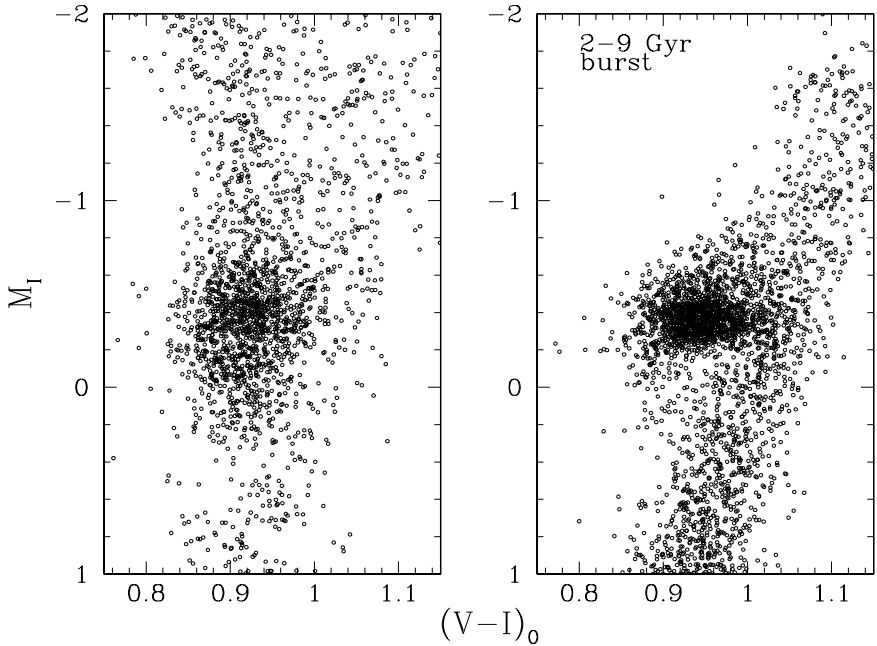,width=3.5in,height=4.8in,angle=-90}
\end{center}
\caption{{\it (Left)} Simulated color-magnitude diagram
for the LMC red clump based on fiducial star formation history 
adopted by Salaris et al.~(2003b).
{\it (Right)} Another simulated color-magnitude diagram, but
for a fictitious burst of field star formation between 2 and 9 Gyrs ago
that produces a redder red clump.
}
\vskip0.3cm
\end{figure}

Finally, 
3 warps in the LMC disk have recently been detected
using a reddening-free color-magnitude index 
$I_0 = I - 1.4\cdot(V-I) + 1.29$
in conjunction with $V$ \& $I$ data for the 
red clump (Olsen \& Salyk 2002;
Subramaniam 2003).  The Olsen \& Salyk (2002) warp in the South-West
quadrant of the disk, and the more significant of the
two warps found by Subramaniam (2003), specifically the warp at the East end 
of the bar, are both coincident with regions of higher reddening.
Olsen \& Salyk (2002) suggested that a foreground dust cloud
was responsible in the South-West, but this does not
explain a second coincidentally
reddened warp at the East end of the bar.
No other distance indicator has yet
confirmed the existence of these warps.

The star formation histories (SFHs) of the 
LMC fields with 
warps have not yet been derived.
Therefore, it is possible that the warps are 
caused by a change in the
mix of young and old red clump stars.
Compare the two simulated
color-magnitude diagrams of the LMC red clump shown in Figure~2
(M. Salaris, {\it priv.~comm.\,}).
The left panel 
corresponds to the fiducial SFH
adopted by Salaris et al.~(2003b), which is constrained
by observations of mostly main sequence stars.
The right panel shows 
an illustrative example of a redder red clump
associated with a fictitious SFH that would be difficult to detect
except for the color of its clump.
A higher fraction of the stars on the right mixed in
with the stars on the left in certain regions of the LMC
could be misinterpreted as physical warps that are coincidentally
also in regions of high reddening.

The LMC has both
young and old populations with distinct properties,
although the detailed nature of the latter is not yet well known.
Therefore, it is possible that 
old clump stars could be mixed
with young clump stars in unexpected ways to
create complex 
color-magnitude variations across the LMC.
New $K$-band data 
would help to test this 
hypothesis~(Alves et al.~2002).

\clearpage

{\bf APPENDIX A }

\begin{footnotesize}

{\it Long-form Abstract for ``Highlights of Astronomy, Vol.~13,''
O. Engvold, ed. }

{\bf A Review of the Distance and Structure of the Large Magellanic Cloud }

The debate about the distance to the 
Large Magellanic Cloud (LMC) has an epic
history full of controversial and dramatic claims
(i.e., see review by A. Walker in ``Stellar Candles for the
Extragalactic Distance Scale,'' astro-ph/0303035),
and yet in recent years a standard distance modulus has emerged
due primarily to the
completion of the {\it Hubble Space Telescope} ({\it HST\,})
key project to measure the Hubble constant
(W. Freedman et al.~2001, ApJ, 553, 47).
The adopted standard distance modulus, 
$\mu_{0} = 18.5 \pm 0.1$ mag,
yields $H_0 = 71\,\pm\,10$
km s$^{-1}$ Mpc$^{-1}$ (total error)
in excellent agreement with that
derived from the {\it Wilkinson Microwave Anisotropy Probe}:
$H_0 = 72\,\pm\,5$ (D. Spergel et al.~2003, ApJS, 148, 175),
which lends considerable support to its accuracy.

The average of 14 recent measurements of the 
distance to the LMC
implies a true modulus of 18.50 $\pm$ 0.02 mag,
and demonstrates a trend 
in the past 2 years of convergence
toward a standard value.
Table~1 is a summary of the distance results by method.
Where both random ($r$) and systematic ($s$) errors are given, I adopt
their arithmetic sum as the total error, 
and I use this to weight the average.  
I exclude one result, and two are assigned
new error bars as noted.  
Last,
I adopt the average distance modulus for the 4 
eclipsing-binary estimates as one result.
Note that
the eclipsing binaries yield a consistent mean distance,
but their scatter is twice as large as their
average measurement error.
The reduced chi-squared 
is about 0.3 for the final average, and
thus the adopted error bars 
are probably too conservative
(the average uncertainty is $\pm$0.08~mag).

The recent results for the LMC distance
demonstrate a remarkably high level
of consistency.
The possibility
that a consensus on the LMC distance 
has been reached seems much more
plausible now than it did 
at the conclusion of the {\it HST} key project only two years ago. 
A great American sports journalist once said famously, 
``The opera ain't over till the fat lady sings,'' 
to make a point that the outcome of a series of games was not yet 
determined.  {\it Regarding the convergence of published LMC distance 
results, I suggest to you that the fat lady has begun to sing.}

There is no standard model for LMC structure, at least not yet,
but in a series of recent papers 
a new understanding has emerged
(R. van der Marel et al.~2002, ApJ, 124, 2639).
It has been established that the photometric major axis 
of the LMC disk is
about 50$^{\circ}$ away from the line-of-nodes
(along which the disk is at a constant distance equal
to the center-of-mass distance),
and thus
the LMC is intrinsically elongated.
Several independent estimates for the center of the LMC 
are consistent with the optical center; the exception is
the center derived from the velocity field of HI gas.
The velocity field of carbon stars 
and simple equilibrium models indicate that
the LMC disk is thick like the Galactic thick disk and flared.
The vertical exponential scale height of carbon stars attains
a value of about 1~kpc at locations 5-6 degrees from center,
and is only about 100 pc in the central bar region.

The LMC is the nearest example besides the Milky Way of a
disk galaxy that has a stellar halo
(D. Minniti et al.~2003, Science, 301, 1508).
Although the LMC has a halo, it does not 
have a bulge, and in this 
way it is similar to M33.
It not known whether or not the Population II distance
indicators like RR Lyrae variables are distributed in the
same way as the young and intermediate-age stellar populations
which represent the bulk of the LMC disk, and hence whether or not
geometric inclination-corrections appropriate for the disk should apply.
However, the kinematics of the halo and disk stars
in the LMC are distinct.

Finally, I note that the
warps in the LMC disk 
recently discovered using the color 
and magnitude of the red clump
(K. Olsen \& C. Salyk, 2002, AJ, 124, 2045)
are suspiciously found in regions of high reddening,
and could be caused by a change in the
mix of young and old red clump stars which would 
bias the reddening correction.
Old and intrinsically redder red clump stars, if they exist,
could be distributed in a different manner
than the young (disk) clump stars, 
and thus it is possible
for the population mix of the clump
to vary in unexpected ways
across the LMC. \\

\end{footnotesize}
\begin{scriptsize}

\noindent
\begin{minipage}{5in}
Table 1. LMC Distance Moduli in Past 2 Years \\
\begin{tabular}{p{1.2in}p{1.6in}p{0.8in}}
\hline
 Method & $\mu_0$ \\
\hline
Red Clump & $18.493\,\pm\,0.033_r\,\pm\,0.03_s$ \\
Red Clump & $18.471\,\pm\,0.008_r\,\pm\,0.045_s$ \\
Red Clump & $18.54\,\pm\,0.10$ \\
Cepheid & 18.48  & $\pm0.1$ \\
Cepheid & $18.55\,\pm\,0.02_r$ & $\pm0.1$ \\
Cepheid & $18.54\,\pm\,0.29$  \\
RR Lyrae & $18.50\,\pm\,0.07$ \\
RR Lyrae & $18.43\,\pm\,0.06_r\,\pm\,0.16_s$ \\
RR Lyrae & $18.55\,\pm\,0.07$ \\
RR Lyrae & $18.48\,\pm\,0.08$ \\
Mira & $18.48\,\pm\,0.08$ \\
Main Sequence & 18.5 --- 18.7 & excl. \\
Main Sequence & $18.58\,\pm\,0.08$ \\
SN 1987A & $18.46\,\pm\,0.12$ \\
\ \ EB (HV\,982) & \ \ $18.51\,\pm\,0.05$ \\
\ \ EB (EROS\,1044) & \ \ $18.38\,\pm\,0.08$  \\
\ \ EB (HV\,5936) & \ \ $18.18\,\pm\,0.09$  \\
\ \ EB (HV\,982) & \ \ $18.63\,\pm\,0.08$  \\
Eclipsing Binaries & $18.46\,\pm\,0.08$ & ave. of 4 \\
{\bf Wtd. Ave.} & {\bf 18.50$\,\pm\,$0.02} & {\bf s.d.= 0.04 } \\
\hline
\hline
\end{tabular}
\end{minipage}

\end{scriptsize}

\end{document}